\documentclass[%
 reprint,
 superscriptaddress,
 numerical,
 showpacs,
 amsmath,amssymb,
 aps,
 longbibliography,
 prl,
 floatfix
]{revtex4-2}

\usepackage{graphicx}

\usepackage[%
  colorlinks=true,
  urlcolor=blue,
  linkcolor=blue,
  citecolor=blue
]{hyperref}

\usepackage[normalem]{ulem}

\newcommand{\ket}[1]{| #1 \rangle}
\newcommand{\bra}[1]{\langle#1 |}

\let\Re\relax
\let\Im\relax

\DeclareMathOperator{\diag}{diag}

\DeclareMathOperator{\Re}{Re}
\DeclareMathOperator{\Im}{Im}

\def\tcm{T.C.M. Group, Cavendish Laboratory, University of Cambridge, J.J. Thomson Avenue, Cambridge, CB3 0HE, UK}
\def\DAMTP{DAMTP, University of Cambridge, Wilberforce Road, Cambridge, CB3 0WA, UK}

\begin{document}

\title{Sachdev-Ye-Kitaev circuits for braiding and charging Majorana zero modes}

\author{Jan Behrends}
\affiliation{\tcm}
\author{Benjamin B\'{e}ri}
\affiliation{\tcm}
\affiliation{\DAMTP}

\begin{abstract}
The Sachdev-Ye-Kitaev (SYK) model is an all-to-all interacting Majorana fermion model for many-body quantum chaos and the holographic correspondence.
Here we construct fermionic all-to-all Floquet quantum circuits of random four-body gates designed to capture key features of SYK dynamics.  
Our circuits can be built using local ingredients in Majorana devices, namely charging-mediated interactions and braiding Majorana zero modes. 
This offers an analog-digital route to SYK quantum simulations that reconciles  all-to-all interactions with the topological protection of Majorana zero modes, a key feature missing in existing proposals for analog SYK simulation. 
We also describe how dynamical, including out-of-time-ordered, correlation functions can be measured in such analog--digital implementations by employing foreseen capabilities in Majorana devices.
\end{abstract}

\maketitle

The Sachdev-Ye-Kitaev (SYK) model is an interaction-only toy model for quantum chaos~\cite{Shenker:2014ct,Maldacena:2016hu,Maldacena:2016gp,Cotler:2017fx}, the gauge-gravity duality~\cite{Kitaev2015}, and non-Fermi liquid behavior~\cite{Sachdev:1993hv,Song:2017hd,Davison:2017hf,BenZion:2018fl,Chowdhury:2018ec}.
Its Majorana fermion ingredients~\cite{Kitaev2015} motivated proposals~\cite{Pikulin:2017js,Chew:2017fn} for realizing SYK physics in the solid state using Majorana zero modes~\cite{Kitaev:2001gb,Fu:2008gu,Lutchyn:2010hp,Oreg:2010gk,Mourik:2012je,Alicea:2012hz,Beenakker:2013jb,Nadj:2014ey,Lutchyn:2018hq}.
(See also Refs.~\onlinecite{Garcia:2017bb,Luo:2019fp,Babbush:2019eh} for approaches via bosonic digital quantum simulations.) 
The SYK model however involves all-to-all interactions~\cite{Sachdev:1993hv,Kitaev2015}, which is hard to reconcile with the exponential localization of topologically protected Majorana zero modes~\cite{Kitaev:2001gb,Fu:2008gu,Lutchyn:2010hp,Oreg:2010gk,Mourik:2012je,Alicea:2012hz,Beenakker:2013jb,Nadj:2014ey,Lutchyn:2018hq}.
Existing proposals, focused on analog quantum simulations of the SYK \emph{Hamiltonian}, hence give up topological protection: they  employ delocalized Majorana modes~\cite{Pikulin:2017js,Chew:2017fn} and achieve an interaction-only system by fine-tuning certain parameters to prevent bilinear terms arising from the overlapping wave functions.

\begin{figure}[b]
 \includegraphics[scale=1]{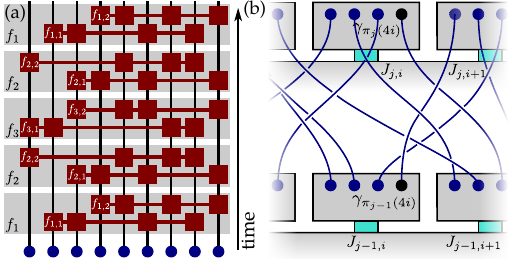}
 \caption{(a)  
 Our model as an all-to-all quantum circuit with four-body gates (maroon) acting on Majorana fermion lines. Shown is a Floquet operator $\mathcal{F}_3$ for $k=9$. 
 (b) Implementing the circuit layers $j-1$ (bottom) and $j$ (top) via superconducting islands (gray rectangles). Each island hosts four Majorana zero modes (blue disks, $\gamma_{\pi_{j-1}(4i)}$ and $\gamma_{\pi_j(4i)}$ in black).
 Charging effects, controlled via coupling (cyan) to a bulk superconductor~\cite{Fu:2010ho,vanHeck:2012bp,Karzig:2017if}, induce the four-body gate $u_i$ [Eq.~\eqref{eq:time_evolution_island}] for island $i$.
Braiding (blue lines), e.g., via ancilla Majoranas~\cite{Sau:2011gj,vanHeck:2012bp} generates the permuted indices $\pi_j (i)$.}
 \label{fig:setup}
\end{figure}

In this work, we focus on the SYK \emph{dynamics} and suggest an approach that reconciles all-to-all interactions with topological protection.
Specifically, we introduce a Majorana fermion model designed to generate interaction-only dynamics respecting SYK model symmetries~\cite{You:2017jj,Behrends:2019jc,Behrends:2020ki}, and which is amenable for analog-digital hybrid quantum simulations using topologically protected Majorana zero modes. 
Our model is a stroboscopic model of many-body quantum chaos~\cite{Yang:2019il,Nahum:2017ef,Nahum:2018bp,Keyserlingk:2018is,Chan:2018im,Chan:2019ba,Chan:2019bb,Chen:2020kd,Kuhlenkamp:2020dl,Sunderhauf:2019hn,Piroli:2020jd,Vijay2020,Nahum:2021ek,Jian:2021jw};
it can be viewed [Fig.~\ref{fig:setup}(a)] as an all-to-all quantum circuit~\cite{Sunderhauf:2019hn,Piroli:2020jd,Vijay2020,Nahum:2021ek,Jian:2021jw}, with Majorana fermion lines~\cite{Bravyi:2002cf,OBrien:2018dx}, defining a Floquet operator~\cite{Chan:2018im,Chan:2019ba,Chan:2019bb}.
Crucially, however, it also arises from local ingredients matching foreseen capabilities of Majorana devices, such as charging-energy-mediated interactions~\cite{Albrecht:2016cw,Vaitiekenas:2020bj} and braid operations [Fig.~\ref{fig:setup}(b)].

\begin{figure}[t]
\includegraphics[scale=1]{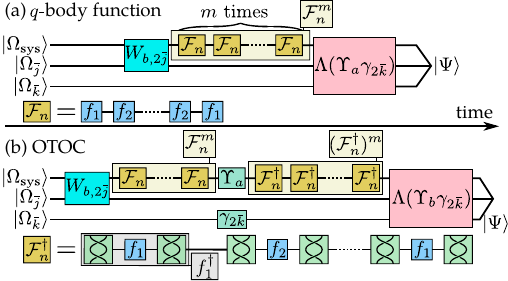}
\caption{Protocol for measuring Majorana monomial correlations with respect to the fermionic vacuum $\ket{\Omega}= \ket{\Omega_\mathrm{sys}} \otimes \ket{\Omega_{\bar{j}}} \otimes \ket{\Omega_{\bar{k}}}$. Here $\ket{\Omega_\mathrm{sys}}$ and $\ket{\Omega_{\bar{j},\bar{k}}}$ are vacua of system and ancilla fermions, respectively.
(a) For $C_{ab}$ in Eq.~\eqref{eq:correlation_bare}, we apply the generalized exchange $W_{b,2\bar{j}}$ (cyan), $m$ Floquet operators $\mathcal{F}_n$ (beige), and $\Lambda (\Upsilon_a \gamma_{2\bar{k}})$ [Eq.~\eqref{eq:fcnot}, pink], on the initial state.
Each $\mathcal{F}_n$ consists of $2n-1$ operators $f_j$ (light blue) for the $j$th circuit layer; cf.\ Eq.~\eqref{eq:floquet_operator}.
(b) For OTOCs, we generate $\mathcal{F}_n^\dagger$ for backward time evolution via double braids (bottom).
}
\label{fig:correlation_protocol}
\end{figure}

We also show how the same capabilities allow one to measure dynamical fermionic SYK correlations (cf.\ Fig.~\ref{fig:correlation_protocol} for a sketch).
These even include out-of-time-ordered correlation functions (OTOCs), often employed to characterize the scrambling of quantum information~\cite{Maldacena:2016gp,Hosur:2016bt,Lantagne:2020it}, but difficult to implement due to OTOCs involving backward time evolution~\cite{Swingle:2016em}.
Our approach achieves a topologically protected evolution reversal using double-braids.
These, together with charge measurements followed by adaptive operations~\cite{Bravyi:2002cf}, allow for an interferometric OTOC protocol~\cite{Swingle:2016em} in our Majorana systems. 
As we will show, one can even probe thermal correlations by applying spectral algorithms~\cite{Abrams:1999jv} and the eigenstate thermalization hypothesis to our Floquet system.

Our model generates the dynamics of Majorana fermions $\gamma_i$, $i=1\ldots k$
via a Floquet operator defined as a $(2n\!-\!1)$-layer [or depth-$(2n\!-\!1)$] all-to-all quantum circuit [Fig.~\ref{fig:setup}(a)].
Each layer generates evolution corresponding to a random partition of $k$ into $\lfloor k/4 \rfloor$ quartets ($\lfloor \dots \rfloor$ is the floor function):
layer $j$ acts via $f_j \equiv \prod_{i=1}^{\lfloor k/4 \rfloor} f_{j,i}$, where $f_{j,i}$ for the $i$th quartet is
\begin{equation}
 f_{j,i} = \exp \left( i J_{j,i} \gamma_{\pi_j (4i-3)} \gamma_{\pi_j (4i-2)} \gamma_{\pi_j(4i-1)} \gamma_{\pi_j(4i)} \right),
 \label{eq:time_evolution_randomized}
\end{equation}
with $J_{j,i}$ a real random variable, and $\pi_j$ a permutation of the $k$ indices. 
[We thus partition by permuting the contiguous partition $(0123)(4567)\ldots$; 
when $k\text{ mod } 4\neq 0$, only the $\gamma_{\pi_j (l)}$ contained in quartets contribute to $f_j$.]
As we shall later explain, to match the antiunitary symmetries of the SYK Hamiltonian time evolution, the Floquet operator must have the structure
\begin{equation}
 \mathcal{F}_n = f_1 f_2 \ldots f_{n-1} f_n f_{n-1} \ldots f_2 f_1.
 \label{eq:floquet_operator}
\end{equation}
Our model can be viewed as a Brownian SYK system~\cite{Sunderhauf:2019hn,Jian:2021jw} built directly from quantum gates (instead of deriving from a time-continuous Hamiltonian) and with a Floquet structure enforced by the antiunitary SYK symmetries.

Turning to the implementation of our model in Majorana devices, we sketch the building blocks in Fig.~\ref{fig:setup}(b):
$4 \lfloor k/4 \rfloor$ of the $k$ Majorana zero modes are distributed over $\lfloor k/4 \rfloor$ superconducting islands, with each island hosting four Majoranas. 
The Majorana wave functions have exponentially small overlap, hence bilinear contributions to the energy can be neglected.
By adjusting the coupling (e.g., via Josephson junctions~\cite{Shnirman:1997iy,Makhlin:2001ba,Koch:2007gz,Schreier:2008gs,Nazarov:2009cm,Fu:2010ho,Xu:2010kb}) of island $i$ to a superconducting reservoir, one can switch between two regimes: (i) strong coupling, where charging effects are absent hence the island Hamiltonian is zero and (ii) weak coupling, where the island Hamiltonian depends only on the island fermion parity with a splitting energy $E_i$~\footnote{See Ref.~\cite{vanHeck:2012bp} for an expression of $E_i$ in terms of characteristic gate voltage, charging and Josephson energy of the island.}. By operating in the weak coupling regime for time $t_i$, island $i$ undergoes time evolution
\begin{equation}
 u_{i} = \exp \left( i J_{i} \gamma_{ 4i-3} \gamma_{4i-2} \gamma_{4i-1} \gamma_{4i} \right),
 \label{eq:time_evolution_island}
\end{equation}
where the dimensionless coupling $J_{i} = t_{i} E_i$.

We generate $\pi_j$  via random braids~\cite{Alicea:2012hz,Yang:2019il}.
A suitable braid operator $R_j$, acting as $ R_j \gamma_{\pi_{j-1} (i)} R_j^\dagger = s_{j,i} \gamma_{\pi_j (i)}$, permutes indices as $\pi_{j-1} (i) \to \pi_j (i)$.
(Here $s_{j,i} = \pm 1$, depending on the details of the braid $R_j$~\footnote{For example, the indices $i$ and $i'$ can be exchanged by either clockwise or anticlockwise braids, resulting in opposite signs.}.)
While $R_j$ is in general nonlocal, it can be built from nearest-neighbor braids, e.g., using the bubble sort algorithm~\cite{OBrien:2018dx}.
[This achieves any $\pi_j$ in at most $O(k^2)$ steps.]
By changing the gate times $t_i \to t_{j,i}$ after each braiding we get the random couplings $J_{j,i}$.
We can thus implement $f_j$ and hence $\mathcal{F}_n$ using Majorana islands.

This implementation uses digital and analog ingredients each to their advantages:
The digital ingredients, i.e., the braids, are topologically protected~\cite{Alicea:2012hz}. 
The $u_i$, in contrast, do not have a topologically protected digital implementation, and even their approximate digital synthesis would come with overhead~\cite{Nielsen:2010ga,Campbell:2017bq,OBrien:2018dx}.
These are thus best suited as analog ingredients. 
Their analog nature notwithstanding, we emphasize that $J_i$ need not match any fine-tuned value; the only requirement to generate $\mathcal{F}_n$ is that the $J_i$ be reproducible. 

Our protocol for measuring dynamical correlations (Fig.~\ref{fig:correlation_protocol}) implements global quenches from certain initial states. 
We focus on initial states of the form $\Upsilon_a \ket{\Omega}$ where $\Upsilon_a $ is a Majorana monomial and $\ket{\Omega}$ is the fermionic vacuum.  
(Our protocol directly generalizes to other initial states; our choice is motivated by capabilities in  Majorana devices.)  
Starting from such product states, we quench the system by evolving with the interaction-only operator $\mathcal{F}_n$.
The resulting overlaps, for example
\begin{equation}
 C_{ab} (m) =  \bra{\Omega} (\mathcal{F}_n^\dagger)^m \Upsilon_a \mathcal{F}_n^m \Upsilon_b \ket{\Omega}
 \label{eq:correlation_bare}
\end{equation}
shown in Fig.~\ref{fig:correlation_protocol}(a), capture correlations between $\Upsilon_b$ at the initial time and $\Upsilon_a$ at the $m$th time step.
The real part of $C_{ab} (m)$ equals the retarded correlation function with respect to $\ket{\Omega}$.
Here we focus on fermion-parity-odd $\Upsilon_{a,b}$ as these carry fundamental fingerprints of SYK-like correlations~\cite{Behrends:2019jc,Behrends:2020ki}.
The OTOC has similar, albeit slightly more complex, structure [Fig.~\ref{fig:correlation_protocol}(b) and Eq.~\eqref{eq:otoc}]. 
To discuss symmetries, we first recall the fermion parity operator $P$~\cite{Kitaev:2001gb} and the antiunitary $T_\pm$~\cite{Behrends:2019jc,Behrends:2020ki}.
Fermion parity, satisfying $P^2=1$, is the even product
\begin{equation}
 P =P^\dagger =
 \begin{cases} i^{k/2} \gamma_1 \gamma_2 \cdots \gamma_k & \text{even $k$} \\
 i^{(k+1)/2} \gamma_1 \gamma_2 \cdots \gamma_k \gamma_\infty & \text{odd $k$,}
 \end{cases}
 \label{eq:parity}
\end{equation}
where for odd $k$, to get an even product, we introduced the Majorana $\gamma_\infty$ at ``infinity"~\cite{Fidkowski:2011dh,Behrends:2019jc}.
($\gamma_\infty$ does not contribute to the dynamics.)
For a set of $k$ Majoranas, there is one or more antiunitary  $T$  such that $T \gamma_{q \neq \infty} T^{-1} = \gamma_{q \neq \infty}$~\cite{deWit:1986gp,Fidkowski:2011dh}.
Following Ref.~\onlinecite{Behrends:2019jc}, we distinguish two possibilities, depending on the interplay with parity: we introduce $T_\pm$ according to $T_\pm P T_\pm^{-1} = \pm P$.
Depending on $k$, one or both of $T_\pm$ are present and satisfy $T_\pm^2 = + 1$ or $T_\pm^2=-1$~\cite{Fidkowski:2011dh,deWit:1986gp,You:2017jj,Behrends:2019jc}.

We can now discuss the symmetries of our model and show that they match those deriving from the SYK Hamiltonian $H_\text{SYK}$. 
Just as $H_\text{SYK}$ (and hence the evolution $U_\text{SYK}=\exp{(-iH_\text{SYK} t)}$ over time $t$), the gates $f_{j,i}$ and hence $\mathcal{F}_n$ conserve fermion parity: $[A,P]=0$ where $A=H_\text{SYK}$, $U_\text{SYK}$, $f_{j,i}$, or $\mathcal{F}_n$. 
This allows the decomposition into blocks with parity eigenvalue $p=\pm1$ as $A = \diag (A^{+}, A^{-})$.
The SYK Hamiltonian (for $b$-body interactions with $b\text{ mod } 4=0$) also has $T_\pm$ as symmetries~\cite{You:2017jj,Behrends:2019jc} and $T_\pm H^{(p)}_\text{SYK} T_\pm^{-1}=H^{(\pm p)}_\text{SYK}$. 
Hence, $T_\pm U_\text{SYK}^{(p)} T_\pm^{-1}=[U_\text{SYK}^{(\pm p)}]^\dagger$. 
To match this, we require
\begin{equation}
T_\pm \mathcal{F}_n^{(p)} T_\pm^{-1} = [ \mathcal{F}_n^{(\pm p)} ]^\dagger
 \label{eq:symmetries_F}.
\end{equation}
The structure introduced in Eq.~\eqref{eq:floquet_operator} achieves precisely this, as can be seen using $T_\pm f_{j,i}^{p} T_\pm^{-1} = [ f_{j,i}^{\pm p}]^\dagger$. 

We next discuss how $T_+$ and $T_-$ influence the spectrum of $\mathcal{F}_n$.
Since $\mathcal{F}_n$ is unitary and $[\mathcal{F}_n,P]=0$, we have
\begin{equation}
 \mathcal{F}_n \ket{\psi_\mu^p} = e^{i \theta_\mu^p} \ket{\psi_\mu^p},
\end{equation}
with real $\theta_\mu^p$ and $P \ket{\psi_\mu^p} = p \ket{\psi_\mu^p}$.
The states $\ket{\psi_\mu^p}$ are also eigenstates of $\mathcal{F}_n^\dagger$ with eigenvalue $e^{-i \theta_\mu^p}$.
When $T_+$ is present, $\ket{\psi_\mu^p}$ and $T_+ \ket{\psi_\mu^p}$ are eigenstates of $\mathcal{F}_n$ with the same eigenvalue and same parity; this impacts the level spacing statistics within each parity block $\mathcal{F}_n^{(p)}$~\cite{You:2017jj}. 
The symmetry $T_-$, conversely, relates opposite parity sectors.
When $T_-$ is present, the states $\ket{\psi_\mu^p}$ and $T_- \ket{\psi_\mu^p}$ both have eigenvalue $e^{i\theta_\mu^p}$, but opposite $p$. 
This symmetry does not affect the level-spacing statistics within $\mathcal{F}_n^{(p)}$, but leads to cross-parity correlations~\cite{Cotler:2017fx,Behrends:2019jc,Behrends:2020ki}.
The features detected by Eq.~\eqref{eq:correlation_bare} with parity-odd $\Upsilon_{a,b}$ include these cross-parity correlations.

We now turn to the correlation functions, starting with $C_{ab}$ in Eq.~\eqref{eq:correlation_bare}.
We define the Majorana monomial
\begin{equation}
 \Upsilon_a = i^{q_a(q_a -1)/2} \gamma_{i_1 (a)} \gamma_{i_2 (a)} \cdots \gamma_{i_{q_a} (a)} 
 \label{eq:def_upsilon}
\end{equation}
as a product of $q_a$ Majorana fermions, where $\{i_j (a) \}$ denotes a set of indices unique for each $a$. 
All $\Upsilon_a$ are Hermitian, unitary, and for odd $q_a$ change fermion parity: $ P \Upsilon_a = (-1)^{q_a} \Upsilon_a P$.
Before discussing the behavior of $C_{ab}$, we briefly comment on how it can be detected via braiding and nondestructive parity measurements.
A central ingredient is a controlled fermionic gate 
\begin{align}
\Lambda(\Upsilon_{a}\gamma_{2\bar{k}})
&= \exp \left( \frac{\pi}{2} \Upsilon_{a} \gamma_{2\bar{k}} \frac{1}{2} (1+i\gamma_{2\bar{j}-1} \gamma_{2\bar{j}} ) \right) \label{eq:fcnot} \\
&= \exp \left( \frac{\pi}{4} \Upsilon_{a} \gamma_{2\bar{k}} \right) \exp \left( i\frac{\pi}{4}\Upsilon_{a}\gamma_{2\bar{k}}\gamma_{2\bar{j}-1} \gamma_{2\bar{j}} \right) , \nonumber
\end{align}
where $\gamma_{2\bar{k}}$, $\gamma_{2\bar{j}-1}$, $\gamma_{2\bar{j}}$ are ancilla Majoranas.
The latter pair defines a control fermionic mode: when their parity $i \gamma_{2\bar{j}-1} \gamma_{2\bar{j}}=1$, the operator $i\Upsilon_{a} \gamma_{2\bar{k}}$ is applied to the target fermionic modes; when $i \gamma_{2\bar{j}-1} \gamma_{2\bar{j}}=-1$, the target modes are left unchanged. 
Because $\Lambda( \Upsilon_{a} \gamma_{2\bar{k}})$ is the product of two $\pi/4$ exponentials, it can be implemented by an adaptive combination of braids and nondestructive parity measurements~\cite{Bravyi:2002cf,OBrien:2018dx}.
The next ingredient is the preparation of the state 
\begin{equation}
 \ket{\Psi} = \Lambda(\Upsilon_{a}\gamma_{2\bar{k}}) \mathcal{F}_n^m W_{b,2\bar{j}} \ket{\Omega} 
 \label{eq:preparation_psi}
\end{equation}
with $W_{a,\bar{j}} = \exp ( \pi \Upsilon_a \gamma_{\bar{j}}/4 )$~\footnote{This operator is a unitary exchange operator for ancilla Majoranas $\gamma_{\bar{j}}$ and parity-odd operators $\Upsilon_a$ since they anticommute.}.
We show the preparation of $\ket{\Psi}$ in Fig.~\ref{fig:correlation_protocol}(a).
A straightforward calculation shows that $C_{ab} (m)$ corresponds to the ancilla averages
\begin{subequations}\begin{align}
 \Re C_{ab} (m) &= i \bra{\Psi} \gamma_{2\bar{k}} \gamma_{2\bar{j}-1} \ket{\Psi} \label{eq:qbody_measurement_real} \\
 \Im C_{ab} (m) &= i \bra{\Psi} \gamma_{2\bar{j}} \gamma_{2\bar{k}}   \ket{\Psi} \label{eq:qbody_measurement_imag} .
\end{align}\label{eq:qbody_measurement}\end{subequations}
The same result also holds for any generic state $\ket{\Omega}$ in Eq.~\eqref{eq:preparation_psi}, so long as $i\gamma_{2\bar{j}-1}\gamma_{2\bar{j}} = 1$.

\begin{figure}
 \includegraphics[scale=1]{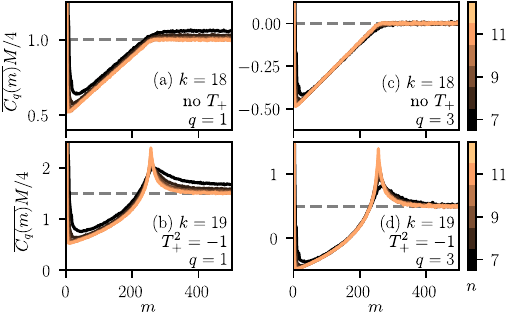}
 \caption{Numerically evaluated $q$-body correlations $C_q (m) = \sum_{a|q_a=q} \Re C_{a,a}(m)/\binom{k}{q}$, rescaled by the Hilbert space dimension $M$ and averaged over up to $2^{16}$ realizations of $J_{j,i}$ taken from a box distribution $J_{j,i} \in [-20,20]$.
 Statistical error bars are smaller than the line width.
 Panels (a)--(b) show $q=1$ and panels (c)--(d) show $q=3$ for $k=18$ (cross-parity correlations set by absence of $T_+$) and $k=19$ Majoranas (cross-parity correlations set by $T_+^2=-1$). The dashed lines show plateau estimates based on random eigenstates subject only to symmetry constraints---the same estimates hold for the plateaus in Hamiltonian SYK evolution~\cite{Behrends:2020ki}. }
 \label{fig:correlations}
\end{figure}

The behavior of $C_{ab}$ shows that the Floquet operator $\mathcal{F}_n$ captures key symmetry-related features of Hamiltonian SYK dynamics. 
In Fig.~\ref{fig:correlations}, we show the numerically evaluated $\Re C_{aa} (m)$ averaged over a large ensemble of couplings $J_{j,i}$, and averaged over all operators $\Upsilon_a$ with fixed $q_a$.
As in Hamiltonian SYK dynamics, we find a ramp and plateau at long times with specific $k$-dependent features:
The ramp connects either smoothly (when $T_+^2 =+1$), with a sharp corner (in absence of $T_+$), or with a kink ($T_+^2=-1$) to a plateau, which reflects the cross-parity level spacing statistics~\cite{Guhr:1998bg,Cotler:2017fx,Behrends:2020ki}. 
The plateau itself reflects the presence and square of $T_-$~\cite{Behrends:2020ki}; 
it toggles between zero and nonzero values with $q_a$ and $k$, with the latter arising only when $(\Upsilon_a T_-)^2=1$~\cite{Behrends:2020ki}.
Since $T_\pm \Upsilon_a T_\pm^{-1} = (-1)^{q_a (q_a-1)/2} \Upsilon_a$, the square $(\Upsilon_a T_-)^2 = (-1)^{n_a} T_-^2$ with $2n_a=q_a -1$.
The operator $\mathcal{F}_n$ even captures quantitative features: 
the nonzero plateau values match excellently those found for Hamiltonian SYK dynamics~\cite{Behrends:2020ki}. 
These signatures, arising already for $n \gtrsim 10$ when $k=18,19$, also show that generating SYK-like chaos needs far fewer four-body gates than the number of interaction terms in the SYK Hamiltonian. 
(We expect $n \sim \log^3 k$ to suffice when the standard deviation $\sigma_{J_{j,i}}\geq 2\pi$~\cite{brown2015decoupling}.)
The operator $\Lambda (\Upsilon_a \gamma_{2\bar{k}})$ also enables us to measure the (unregularized~\footnote{Two types of OTOCs are commonly used~\cite{Liao:2018gw,Romero:2019gf,Lantagne:2020it}: the regularized variant that employs multiple instances of $\rho^{1/4} = \exp ( -\beta H/4)$ for thermal ensembles and the unregularized variant used here.}) 
OTOC with respect to any state $\ket{\Omega}$ (so long as $i\gamma_{2\bar{j}-1}\gamma_{2\bar{j}} = 1$),
\begin{equation}
 F_{ab} (m) = \bra{\Omega} \Upsilon_a (m)  \Upsilon_b (0)  \Upsilon_a (m)  \Upsilon_b (0)\ket{\Omega},
 \label{eq:otoc}
\end{equation}
where $\Upsilon_a (m)=(\mathcal{F}_n^\dagger)^m \Upsilon_a \mathcal{F}_n^m$.
Measuring $F_{ab} (m)$ requires backward time evolution.
Crucially, in our approach this is possible \emph{without} reversing the energy splittings (i.e., without sending $J_{j,i} \to -J_{j,i}$).
Instead, we backward evolve via the double braid $ R_{ii'}^2 = \gamma_{i} \gamma_{i'}$: this sends~\cite{Martin:2020ed}
\begin{align}
 R_{ii'}^2 \gamma_{i } (R_{ii'}^\dagger)^2 = - \gamma_i ,& &
 R_{ii'}^2 \gamma_{i'} (R_{ii'}^\dagger)^2 = - \gamma_{i'} ,
\end{align}
but leaves all other Majoranas invariant. 
Due to
\begin{equation}
 R_{\pi_j (4i),\pi_j (4i+1)}^2 f_{j,i} f_{j,i+1}  (R_{\pi_j (4i),\pi_j (4i+1)}^\dagger )^2
  = f_{j,i}^\dagger f_{j,i+1}^\dagger ,
\end{equation}
this double braid effectively reverses time.
To reverse the whole system's time evolution, we apply double braids to all pairs of islands~\footnote{For an odd number of islands, we require an ancilla Majorana $\gamma_{2\bar{j}}$ and employ $R_{\pi_j (4 \lfloor k/4 \rfloor,2\bar{j})}$.}, giving
\begin{align}
D_j f_j D_j^\dagger = f_j^\dagger ,
& & D_j = \prod_{i=1}^{\lfloor k/8 \rfloor} R_{\pi_j (8i-4),\pi_j (8i-3)}^2 .
\label{eq:double_braid}
\end{align}
The operator
\begin{equation}
 \mathcal{F}_n^\dagger
 = D_1 f_1 D_1^\dagger D_2 f_2 D_2^\dagger \ldots D_n f_n D_n^\dagger \ldots D_2 f_2 D_2^\dagger D_1 f_1 D_1^\dagger
\end{equation}
generates the evolution reverse to $\mathcal{F}_n$.

To evaluate the OTOC, we prepare the state
\begin{equation}
\ket{ \Psi } = \Lambda(\Upsilon_{b}\gamma_{2\bar{k}}) (\mathcal{F}_n^\dagger)^m \Upsilon_{a} \gamma_{2\bar{k}} \mathcal{F}_n^m W_{b,2 \bar{j}}
\ket{ \Omega } ,
\label{eq:psi_otoc}
\end{equation}
as shown in Fig.~\ref{fig:correlation_protocol}(b).
Apart from the ingredients for Eq.~\eqref{eq:correlation_bare}, this requires only the application of $\Upsilon_a \gamma_{2 \bar{k}}$, which is the square of a $\pi/4$ exponential, hence also implementable by braids and nondestructive parity measurements. 
Using this state, the real and imaginary part of the OTOC can be measured using 
\begin{subequations}\begin{align}
 \Re F_{ab} (m) &= i \bra{\Psi} \gamma_{2\bar{j}-1} \gamma_{2\bar{k}} \ket{\Psi} \label{eq:otoc_measurement_real},\\
 \Im F_{ab} (m) &= i \bra{\Psi} \gamma_{2\bar{k}  } \gamma_{2\bar{j}} \ket{\Psi} \label{eq:otoc_measurement_imag}.
\end{align}\label{eq:otoc_measurement}\end{subequations}

\begin{figure}
 \includegraphics[scale=0.9]{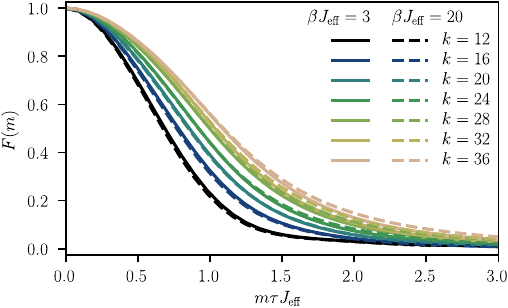}
 \caption{Numerically evaluated single-Majorana OTOC $F = \sum_{i \neq i'} \Re F_{ii'}/(k(k-1))$ averaged over up to $2^{11}$ realizations of box-distributed $J_{j,i} \in [-0.01,0.01]$. 
 The number $k$ of Majoranas is shown by the colors. 
 All curves are for $n=35$; the OTOCs do not appreciably change upon increasing $n$ further.
Solid (dashed) lines show $\beta J_\mathrm{eff} = 3$ ($\beta J_\mathrm{eff} =20$); the statistical error bars are smaller than the line width.
}
 \label{fig:otoc}
\end{figure}

A key aspect of SYK dynamics are holographic features in \emph{thermal} OTOCs.
To study such OTOCs in our dynamical model, we must infuse thermal physics into our Floquet system. 
To this end, we focus on $\mathcal{F}_n$ with $|\theta_\mu| < \pi/2$ for all $\mu$; in this case we can interpret $\theta_\mu=\tau \varepsilon_\mu$ as an energy $\varepsilon_\mu$ times the Floquet period $\tau$ and define a thermal ensemble.
Upon invoking the eigenstate thermalization hypothesis~\cite{Deutsch:2018fy,Sonner:2017bw,ETHfn} for this ensemble, one can probe thermal features by using suitable (approximate) eigenstates of $\mathcal{F}_n$ for $\ket{ \Omega }$. 
Such eigenstates can be prepared using a variant of quantum phase estimation~\cite{Kitaev1995,Abrams:1999jv}, while matching the corresponding $\theta_\mu$ to $\beta/\tau$ (with inverse temperature $\beta$) can be done via analytical results on the SYK density of states~\cite{Garcia:2017dv} (which also describe the distribution of $\theta_\mu$ to a good approximation).

In Fig.~\ref{fig:otoc}, we show OTOC numerics based on this matching procedure. 
We focus on short times, take $\Upsilon_{a,b} \to \gamma_{i,i'}$, and show results for $\beta J_\mathrm{eff} =3,20$.
Here $J_\mathrm{eff}$ is the coupling strength of a (fictitious) SYK Hamiltonian, $\tau J_\mathrm{eff} = \sigma_{\theta} \sqrt{ 4 k^2 (k-4)! /(k-1)!}$~\cite{Garcia:2017dv,Behrends:2019jc}, where $\sigma^2_{\theta}= (4n-3) \lfloor k/4 \rfloor \sigma^2_{J_{j,i}} + O(\sigma^4_{J_{j,i}})$ for $\sigma_{J_{j,i}}\ll 2\pi$~\cite{sigmathetafn}. 
(We keep $\sigma^2_{\theta}\ll \pi$ so that a link to thermal physics is possible; a comparison to Ref.~\onlinecite{Campbell19} suggests that this remains consistent with chaos even for $k\gg 1$ and for $n$ increasing with $k$ to reach convergence.) 
The spacing between OTOCs for different system sizes increases with $\beta$ for $k\geq 20$,
suggesting a Lyapunov exponent decreasing with lowering temperature~\cite{Kobrin:2021bx}. 
Both this, and the faster (slower) decay for larger $\beta$ when $k<20$ ($k>20$) resemble closely the OTOCs in the SYK model~\cite{Shenker:2014ct,Maldacena:2016hu,Maldacena:2016gp,Cotler:2017fx,Kitaev2015,Kobrin:2021bx}.

In this work, we introduced Floquet quantum circuits for generating quantum chaotic dynamics akin to the SYK model.
Our circuits capture key dynamical SYK features, as we have demonstrated via ramps and plateaus in $q$-body correlations $C_{ab}$ and via low-temperature OTOCs.
While we focused on four-body circuits, our model directly generalizes to one with $b$-body gates; there, owing to the same symmetries, we expect analogous features to arise whenever $b \text{ mod } 4=0$. 

We have also suggested an analog-digital hybrid implementation for our model and the measurement of $C_{ab}$ and the OTOCs. 
The digital components are braiding and state preparation,  the analog components are the gates $f_{j,i}$ for evolving Majorana quartets. 
This hybrid approach consists only of well-known Majorana operations and combines the best of the analog and digital worlds: its digital parts enjoy topological protection, while its analog gates appear where digital execution would incur considerable overhead (e.g., due to magic state distillation and gate synthesis~\cite{Nielsen:2010ga,Campbell:2017bq,OBrien:2018dx}).
Beyond these, a native fermionic implementation (instead of qubit-based bosonic simulation) of a fermionic model has further advantages due to the matching locality properties of simulation and model ingredients~\cite{Bravyi:2002cf,OBrien:2018dx}.

Our SYK quantum circuits suggest several further directions, including studying local forms of the model or effects leading to nonunitary SYK quantum circuits.
Such nonunitary circuits should not only have analog-digital hybrid implementations, but may also arise naturally from perturbations to unitary SYK circuits, which makes exploring their physics especially interesting.

\begin{acknowledgments}
This work was supported by the ERC Starting Grant No.\ 678795 TopInSy and the EPSRC grant EP/S019324/1.
\end{acknowledgments}

\bibliography{dyn_syk}

\begin{thebibliography}{81}
\bibitem [{ETH()}]{ETHfn}%
  \BibitemOpen
\bibfield  {number} {  }\href@noop {} {}\bibinfo {note} {{We verified for a
  range of $\beta/\tau$ values and system sizes that eigenstates with
  $\theta_\mu$ matching the ensemble's mean energy reproduce thermal two- and
  four-point correlations }}\BibitemShut {NoStop}%
\bibitem [{sig()}]{sigmathetafn}%
  \BibitemOpen
  \href@noop {} {}\bibinfo {note} {{Including the $O( \sigma_{J_{j,i}}^4)$
  term, we find $\sigma^2_{\theta}= (4n-3) \lfloor k/4 \rfloor
  \sigma_{J_{j,i}}^2 - (2/3) (1 - \eta) (2n^2 -5 n +3 ) \lfloor k/4 \rfloor^2
  \sigma_{J_{j,i}}^4 + O (\sigma_{J_{j,i}}^6)$ where, as in
  Ref.~\onlinecite{Garcia:2017dv}, $\eta = [1/{\binom{k}{4}}] \sum_{r=0}^4
  (-1)^r \binom{4}{r} \binom{k-4}{4-r}$. We use this $\sigma^2_{\theta}$
  expression in $J_\mathrm{eff}$ for the numerical OTOC simulations
  }}\BibitemShut {NoStop}%
\end{thebibliography}

\end{document}